\documentclass[journal]{IEEEtran}
\usepackage{graphics}
\usepackage{mathrsfs}
\usepackage{color}
\usepackage{amsfonts} 
\usepackage{color} 
\usepackage{float}
\usepackage{cite}
\usepackage{graphicx}
\usepackage{amsmath}
\usepackage{amsthm}
\usepackage{multicol}
\usepackage{multirow}
\usepackage{color}
\usepackage{ifpdf}
\usepackage{flushend}
\usepackage{txfonts}

\usepackage{CJK}

\theoremstyle{definition} \newtheorem{remark}{Remark}
\theoremstyle{definition} \newtheorem{proposition}{Proposition}
\theoremstyle{definition} \newtheorem{theorem}{Theorem}
\theoremstyle{definition} \newtheorem{lemma}{Lemma}
\theoremstyle{assumption} 
\theoremstyle{definition} \newtheorem{definition}{Definition}
\theoremstyle{definition} 
\theoremstyle{definition} 

\usepackage{ifpdf}
\ifpdf
  \usepackage{epstopdf}
\fi

\newcommand{\R}{\mathbb{R}}

\newcommand{\K}{\mathcal{K}}

\newcommand{\setS}{\mathcal{S}}

\begin{document}
\begin{CJK}{UTF8}{gbsn}

\title{Small-Gain Theorem for Safety Verification of Interconnected Systems}

\author{Ziliang~Lyu,~
        Xiangru~Xu,~
        and~Yiguang~Hong

\thanks{Z. Lyu (e-mail: ziliang\_lyu@outlook.com) and Y. Hong (e-mail: yghong@iss.ac.cn) are with the Department of Control Science and Engineering, Tongji University, Shanghai, China}

\thanks{X. Xu (e-mail: xiangru.xu@wisc.edu) is with the Department of Mechanical Engineering, University of Wisconsin-Madison, Madison, WI, USA.}
}

\maketitle

\begin{abstract}
A small-gain theorem in the formulation of barrier function is developed in this work for safety verification of interconnected systems. This result is helpful to verify input-to-state safety (ISSf) of the overall system from the safety information encoded in the subsystem's ISSf-barrier function. Also, it can be used to obtain a safety set in a higher dimensional space from the safety sets in two lower dimensional spaces.


\end{abstract}


\section{Introduction}

Safety is crucial in the design of control systems, which requires any trajectories of the control systems initialized in a prescribed safety set to be kept out of the unsafe region. Applications concerned with safety are ubiquitous in our daily life, ranging from the cars on the ground to the aeroplanes in the air. For example, autonomous vehicles are equipped with lane keeping modules \cite{xu2017correctness}, a robot team is designed to avoid the collision between robots \cite{wang2017safety}, and aircrafts must satisfy the safety requirement during takeoff and landing \cite{balachandran2016verification}.

For safety-critical systems, a rigorous verification of safety is the first step towards other control objectives (e.g., stabilization and regulation). In general, the techniques for safety verification can be classified into two sorts: model checking \cite{clarke2018model} and deductive verification \cite{manna2012temporal}. Compared with the former one, deductive verification provides a safety certificate by mathematical inferences rather than exhaustively checking all of the possible system behaviors. The barrier function gives a promising deductive verification approach, with a set of Lyapunov-like criterions for safety verification; e.g., see \cite{prajna2007framework, ames2014control, xu2015robustness, ames2016control, kolathaya2018input, wieland2007constructive, coogan2014dissipativity, romdlony2016new} for more details. In \cite{prajna2007framework}, barrier functions were used to formulate the verification tasks as convex programming problems. A framework by combining control Lyapunov functions and barrier functions was proposed in \cite{ames2014control} to balance the objective of stabilization and safety verification. In \cite{xu2015robustness}, a novel barrier function, called the zeroing barrier function, was proposed for establishing safety and analyzing the robustness of safe sets. Later, \cite{kolathaya2018input} redefined a new notion of input-to-state safety (ISSf) and provided a sufficient condition in the sense of barrier function to check ISSf, which could be regarded as the counterpart of Sontag's input-to-state stability (ISS) \cite{sontag1995characterizations} for safety verification. However, it should be mentioned that most of these results were not given for interconnected systems.

The small gain technique is effective for interconnected systems to keep good properties, which completes the verification tasks of a large-scale system to be completed by analyzing its less complicated subsystems. In the stability analysis, this technique has been investigated \cite{jiang1996a, Jiang1994Small, yang2015lyapunov, long2013small, Nesic2008A}. For example, the Lyapunov-based small-gain theorem developed in \cite{Jiang1994Small} has been found useful to establish ISS for interconnected systems. This result has been extended to switched systems \cite{yang2015lyapunov} and hybrid systems \cite{Nesic2008A} in recent years.

Considering that the barrier function provides a Lyapunov-like verification approach, it is natural to ask a question: whether there exists a small-gain condition such that one can establish ISSf for the overall system from the safety information encoded in the subsystem's ISSf-barrier function. This question is not trivial. First, the existing small-gain theorems for stability rely on some nonnegative assumptions, which cannot be met in safety verification. Therefore, we cannot directly employ the tools developed in \cite{jiang1996a, Jiang1994Small, yang2015lyapunov, long2013small, Nesic2008A} to analyze the small-gain condition for safety verification. Second, the ISS plays important roles on the small-gain conditions, but there are no small-gain results of ISSf.

The objective of this paper is to develop a small-gain theorem based on barrier functions for safety verification of interconnected system, in order to establish a higher dimensional safety set from two lower dimensional safety sets. Based on the proposed ISSf-barrier function, we prove that, if the absolute value of the composition of ISSf gains is smaller than that of the identity function, then the overall system is ISSf. This condition is different from traditional small-gain conditions for stability.

\textbf{Notations and Terminologies.} Throughout this paper, `$\circ$' denotes the composition operator, i.e., $f\circ{g}(s)=f(g(s))$; `T' denotes the transpose operator; $|\cdot|$ denotes the Euclidean norm; $\R$ and $\R_{\geq0}$ denote the set of real numbers and nonnegative real numbers, respectively. For any measurable  function $u:\R_{\geq0}\rightarrow\R^m$, $\|u\|=\sup\{|u(t)|,t\geq0\}$. In the following, we review the comparison functions, whose details can be founded in \cite{xu2015robustness} and \cite{khalil2002nonlinear}. A continuous function $\gamma$: $\mathbb{R}_{\geq0}\rightarrow\mathbb{R}_{\geq0}$ with $\gamma(0)=0$ is of class $\mathcal{K}$, if it is strictly increasing. Moreover, a class $\K$ function $\gamma$ is of class $\mathcal{K}_\infty$ if it is unbounded. A continuous function  $\gamma:\R\rightarrow\R$ with $\gamma(0)=0$ is of extended class $\K$ if it is strictly increasing. In particular, an extended class $K$ function $\gamma$ is of extended class $\K_\infty$ if it is unbounded.

\section{Preliminaries}
Consider the system
\begin{flalign}
    \dot{x}=f(x,u),\;\;x(t_0)=x_0
    \label{eq:single-system}
\end{flalign}
where $x\in\R^n$ is the state, $u\in\R^m$ is the external input, and the vector field $f:\R^n\times\R^m\rightarrow\R^n$ is locally Lipschitz continuous with $f(0,0)=0$.

We first review two definitions related to safety.

\begin{definition}[Forward Robust Invariance \cite{blanchini2008set}]
A set $\setS$ is forward robustly invariant if for every $x_0\in\setS$, there exists an external input $u$ such that the solution $x(t)$ to (\ref{eq:single-system}) satisfies $x(t)\in\setS$ for all $t\geq{t}_0$.
\end{definition}

\begin{definition}[Input-to-State Safety \cite{kolathaya2018input}]
The system (\ref{eq:single-system}) is input-to-state safe (ISSf) on the set
\begin{flalign}\label{eq:ISSfset}
        \setS_{\gamma(\|u\|)}=\{x\in\R^n:h(x)+\gamma(\|u\|)\geq0\}
\end{flalign}
with respect to the external input $u$, if $\setS_{\gamma(\|u\|)}$ is forward robustly invariant. Herein, $h:\R^n\rightarrow\R$  is a continuously differentiable function, and $\gamma$ is a function of class $\K_\infty$. In particular, we say that system (\ref{eq:single-system}) is safe if there is no external input (i.e., $u=0$), in which case $S_{\gamma(\|u\|)}=\{x\in\R^n:h(x)\geq0\}$.
\end{definition}

Because in the proof of our small-gain theorem we need to handle nonsmooth functions, here we introduce the Dini derivative and corresponding results.

\begin{definition}[Dini Derivative \cite{friedrichs1985advanced}]
Consider a function $h:\R\rightarrow\R$. The upper and the lower Dini derivatives of $h$ at $r\in\R$ are, respectively, defined by
\begin{flalign}
    &D^+h(r)=\limsup_{s\rightarrow0^+}\frac{h(r+s)-h(r)}{s},
     \nonumber\\
    &D_+h(r)=\liminf_{s\rightarrow0^+}\frac{h(r+s)-h(r)}{s}.
     \nonumber
\end{flalign}
\end{definition}

Given a continuous function $h:\R\rightarrow\R$, it is known that $h$ is not increasing at $r\in\R$ if and only if the Dini derivative of $h$ at $r$ is nonpositive, and is not decreasing if and only if the Dini derivative of $h$ at $r$ is nonnegative.

\begin{lemma}\label{lemma:dini}
For $i=1,2$, suppose that $h_i:\R^{n_i}\rightarrow\R$ is locally Lipschitz and let $h(x)=\min\{h_1(x_1),h_2(x_2)\}$. Then $h(x)$ is locally Lipschitz as well, and
\begin{flalign}
    &D_+{h}(x)=\dot{h}_1(x_1),
     \;\;\;\;\;\;\;\;\;\;\;\;\;\;\;\;\;\;\;\;\,\text{if } h_1(x_1)<h_2(x_2);
     \nonumber\\
    &D_+h(x)=\dot{h}_2(x_2),
     \;\;\;\;\;\;\;\;\;\;\;\;\;\;\;\;\;\;\;\;\,\text{if } h_1(x_1)>h_2(x_2);
     \nonumber\\
    &D_+h(x)=\min\{\dot{h}_1(x_1),\dot{h}_2(x_2)\},
     \;\;\text{if } h_1(x_1)=h_2(x_2).
     \nonumber
\end{flalign}
\end{lemma}

Let $f_i(x)=-h_i(x)$ and $f(x)=-h(x)$. Then Lemma \ref{lemma:dini} follows directly from \cite[Theorem 2.1]{clarke1975generalized} due to  $D^+f(x)=-D_+h(x)$ and $\max\{\dot{f}_1(x_1),\dot{f}_2(x_2)\}=-\min\{\dot{h}_1(x_1),\dot{h}_2(x_2)\}$.

\begin{lemma}[Comparison Principle \cite{rouche1977stability}]\label{lem:compar}
Consider
\begin{flalign}
    \dot{x}=f(x),\;\;x(t_0)=x_0
    \nonumber
\end{flalign}
where $f:\R\rightarrow\R$ is locally Lipschitz. Let $[t_0,T)$ be the maximal interval of existence of the solution $x(t)$. Suppose that $y(t)$ is a continuous function with
\begin{flalign}
    D_+y(t)\geq{f}(y(t)),\;\;y(t_0)\geq{x}_0.
\end{flalign}
Then $y(t)\geq{x}(t)$ for all $t\in[t_0,T)$.
\end{lemma}

\section{ISSf-Barrier Function}

In this section, we present a new notion of ISSf-barrier function, which is suitable to characterize the ISSf gains of our small-gain condition.

\begin{definition}[ISSf-Barrier Function]
A continuously differentiable function $h:\R^n\rightarrow\R$ is called an ISSf-barrier function for system (\ref{eq:single-system}), if there exist functions $\gamma$ of class $\K$ and $\alpha$ of extended class $\K$ such that for each $u\in\R^m$,
\begin{flalign}\label{eq:ISSf-BF-def}
    h(x)\leq-\gamma(|u|)\Rightarrow\nabla{h}(x)f(x,u)\geq-\alpha(h(x)).
\end{flalign}
Herein, $\gamma$ is referred to as the ISSf gain.
\end{definition}

Note that \cite{kolathaya2018input} has presented another definition of ISSf-barrier functions; referring to (\ref{eq:ISSf-BF-def-dissipation}) below. The relation between the ISSf-barrier function defined in (\ref{eq:ISSf-BF-def}) and the one of \cite{kolathaya2018input} is shown by the following result.

\begin{proposition}
A continuously differentiable function $h:\R^n\rightarrow\R$ is an ISSf-barrier function for system (\ref{eq:single-system}) with $\alpha$ of extended class $\K_\infty$ if and only if there exist functions $\chi$ of extended class $\K_\infty$ and $\phi$ of class $\K_\infty$ such that
\begin{flalign}\label{eq:ISSf-BF-def-dissipation}
   \nabla{h}(x)f(x,u)\geq-\chi(h(x))-\phi(|u|).
\end{flalign}
\end{proposition}

\noindent
\textbf{Proof.}
Select a constant $c\in(0,1)$. Then it follows from (\ref{eq:ISSf-BF-def-dissipation}) that
\begin{flalign}
    h(x)\leq-\hat{\phi}(|u|)\Rightarrow\nabla{h}(x)f(x,u)\geq-(1-c)\chi(h(x))
    \nonumber
\end{flalign}
where $\hat{\phi}(r)=-\chi^{-1}(-\phi(r)/c)$. The remainder is to verify that $\hat{\phi}$ is of class $\K$. To see this, note that $\hat{\chi}(r)=-\chi^{-1}(-r/c)$ is continuous and strictly increasing on the interval $[0,\infty)$, which in turn, implies that $\hat{\phi}(r)=\hat{\chi}\circ\phi(r)$ is of class $\K$.

Then we show that the converse is also true. Assume that (\ref{eq:ISSf-BF-def}) holds with some $\alpha$ of extended class $\K_\infty$ and $\gamma$ of class $\K$. According to (\ref{eq:ISSf-BF-def}), we have the following two cases.

\textbf{Case 1.} $h(x)\leq-\gamma(|u|)$. In this case, there exists a function $\phi$ of class $\K_\infty$ such that,
\begin{flalign}
    \nabla{h}(x)f(x,u)\geq-\alpha(h(x))\geq-\alpha(h(x))-\phi(|u|).
\end{flalign}

\textbf{Case 2.} $h(x)>-\gamma(|u|)$. Since $\alpha$ is of extended class $\K_\infty$,
\begin{flalign}
    \nabla{h}(x)f(x,u)+\alpha(h(x))
    &\geq\nabla{h}(x)f(x,u)+\alpha(-\gamma(|u|))
     \nonumber\\
    &\geq\inf_{h(x)\geq-\gamma(|u|)}\nabla{h}(x)f(x,u)+\alpha(-\gamma(|u|))
    \label{eq:idontknown}
\end{flalign}
Let $\tilde{\phi}(r)=\max\{0,\hat{\phi}(r)\}$ with
\begin{flalign}
    \hat{\phi}(r)=-\inf_{h(x)\geq-\gamma(|r|)}\nabla{h}(x)f(x,r)+\alpha(-\gamma(|r|)).
    \nonumber
\end{flalign}
This implies that $\tilde{\phi}$ is continuous and $\tilde{\phi}(0)=0$ (recalling $f(0,0)=0$). Note that $\tilde{\phi}(r)\geq0$ for each $r\geq0$. By combining this with (\ref{eq:idontknown}), we can select a class $\K_\infty$ function $\phi$ with $\phi(r)\geq\tilde{\phi}(r)$ for all ${r}\geq0$ such that
\begin{flalign}
    \nabla{h}(x)f(x,u)+\alpha(h(x))\geq-\phi(|u|).
    \nonumber
\end{flalign}

By combining the two cases above, we conclude that the converse is true as well.
\hfill $\Box$\vskip 5pt

Then we are ready to present the main result of this section, which can be used as a tool to establish ISSf.

\begin{theorem}\label{thm:ISSf}
Consider the system (\ref{eq:single-system}) and the set $\setS_{\gamma(\|u\|)}$ defined in (\ref{eq:ISSfset}). Suppose that $h:\R^n\rightarrow\R$ is an ISSf-barrier function satisfying (\ref{eq:ISSf-BF-def}). Then,

(i) $\setS_{\gamma(\|u\|)}$ is forward robustly invariant, and system (\ref{eq:single-system}) is ISSf on $\setS_{\gamma(\|u\|)}$;

(ii) the safety set $\setS_{\gamma(\|u\|)}$ is ISS.
\end{theorem}

\noindent
\textbf{Proof.} We show the conclusion of (i) by contradiction. If this is not true, then there exist some $t\geq{t_0}$ and some $\epsilon>0$ such that $h(x(t))<-\gamma(\|u\|)-\epsilon$. Let $\tau=\inf\{t\geq{t_0}:h(x(t))\leq-\gamma(\|u\|)-\epsilon\}$. Therefore, $h(x(\tau))\leq-\gamma(\|u\|)\leq-\gamma(|u(\tau)|)$, which in turn, implies that
\begin{flalign}
    \dfrac{\text{d}}{\text{d}t}\bigg|_{t=\tau}h(x(t))
    &=\nabla{h}(x(\tau))f(x(\tau),u(\tau))
     \nonumber\\
    &\geq-\alpha(h(x(\tau)))>0.
\end{flalign}
Thus, $h(x(t))\leq{h}(x(\tau))$ for some $t$ in $(t_0,\tau)$. This contradicts the minimality of $\tau$ and consequently, $x(t)\in\setS_{\gamma(\|u\|)}$ for all $t\geq{t_0}$.

Then we consider (ii). Construct a Lyapunov-like function as in \cite{xu2015robustness}:
\begin{flalign}
    V(x)=\left\{
           \begin{array}{ll}
             0, & \text{if } x\in\setS_{\gamma(\|u\|)} \\
             -h(x), & \text{if } x\in\R^n\setminus\setS_{\gamma(\|u\|)}
           \end{array}
         \right.
    \nonumber
\end{flalign}
With (\ref{eq:ISSf-BF-def}), we have
\begin{flalign}
     V(x)\geq\gamma(|u|)\Rightarrow\nabla{V}(x)f(x,u)\leq\alpha(-V(x)),\;\;\forall{x}\in\R^n\setminus\setS_{\gamma(\|u\|)}.
     \nonumber
\end{flalign}
Moreover, with \cite[Lemma 2.14]{sontag1995characterizations} and \cite[Proposition 4]{xu2015robustness}, we conclude that the set $\setS_{\gamma(\|u\|)}$ is ISS.
\hfill $\Box$

%
%
%
%

\section{Small-Gain Theorem for Safety Verification}
Consider the interconnected system
\begin{subequations}\label{eq:interconnect-system}
\begin{flalign}
    &\dot{x}_1=f_1(x_1,x_2,u_1)
     \\
    &\dot{x}_2=f_2(x_1,x_2,u_2)
\end{flalign}
\end{subequations}
where, for $i=1,2$, $x_i\in\R^{n_i}$, $u_i\in\R^{m_i}$ and $f_i:\R^{n_1}\times\R^{n_2}\times\R^{m_i}\rightarrow\R^{n_i}$ is locally Lipschitz. Let $x=(x_1^\text{T},x_2^\text{T})^\text{T}$, $u=(u_1^\text{T},u_2^\text{T})^\text{T}$ and $f(x,u)=(f_1(x_1,x_2,u_1)^\text{T},f_2(x_1,x_2,u_2)^\text{T})^\text{T}$

The small-gain theorem for safety verification of the interconnected system (\ref{eq:interconnect-system}) is summarized as follows.

\begin{theorem}\label{thm:small-gain}
Consider the interconnected system given in (\ref{eq:interconnect-system}). Assume that there exist functions $\alpha_i$ of extended class $\K$, $\phi_i$ of extended class $\K_\infty$ and $\gamma_i$ of class $\K$ such that
\begin{flalign}
    &h_1(x_1)\leq-\max\{\phi_1(|h_2(x_2)|),\gamma_1(|u_1|)\}
     \nonumber\\
    &\;\;\;\;\;\;\;\;\;
     \Rightarrow\nabla{h}_1(x_1)f_1(x_1,x_2,u_1)\geq-\alpha_1(h_1(x_1)),
     \label{eq:barrier-type-conditionA}\\
    &h_2(x_2)\leq-\max\{\phi_2(|h_1(x_1)|),\gamma_2(|u_2|)\}
     \nonumber\\
    &\;\;\;\;\;\;\;\;\;
    \Rightarrow\nabla{h}_2({x}_2)f_2(x_1,x_2,u_2)\geq-\alpha_2(h_2(x_2)).
     \label{eq:barrier-type-conditionB}
\end{flalign}
Then there exist an ISS-barrier function $h:\R^{n_1}\times\R^{n_2}\rightarrow\R$, which is dependent on $h_1$ and $h_2$, and a class $\K_\infty$ function $\gamma:\R_{\geq0}\rightarrow\R_{\geq0}$, which is dependent on $\gamma_1$ and $\gamma_2$, such that the interconnected system (\ref{eq:interconnect-system}) is ISSf on the set
\begin{flalign}
    \setS_{\gamma(\|u\|)}
    =\{x\in\R^n:h(x)+\gamma(\|u\|)\geq0\}
     \label{eq:inter-safe-set}
\end{flalign}
with respect to the external input $u$, if
\begin{flalign}
    \left\{
      \begin{array}{ll}
        \phi_1\circ\phi_2(r)>r, & \forall{r}<0; \\
        \phi_1\circ\phi_2(r)<r, & \forall{r}>0.
      \end{array}
    \right.
    \label{eq:small-gain}
\end{flalign}
\end{theorem}

\begin{remark}\label{remark:small-set}
Clearly, $h_i(x_i)$ is an ISSf-barrier function for $x_i$-subsystem. According to Theorem \ref{thm:ISSf}, the barrier-type conditions given in (\ref{eq:barrier-type-conditionA}) and (\ref{eq:barrier-type-conditionB}) imply that the sets
\begin{flalign}
    &\setS_{1}=\big\{x_1:h_1(x_1)+\max\{\phi_1(\|h_2(x_2)\|),\gamma_1(\|u_1\|)\}\geq0\big\},
     \nonumber\\
    &\setS_{2}=\big\{x_2:h_2(x_2)+\max\{\phi_2(\|h_1(x_1)\|),\gamma_2(\|u_2\|)\}\geq0\big\}
     \nonumber
\end{flalign}
are forward robustly invariant. Intuitively, Theorem \ref{thm:small-gain} can be regarded as a tool to establish a robustly invariant set $\setS_{\gamma(\|u\|)}$ in a higher dimensional space from the robustly invariant sets $\setS_{1}$ and $\setS_{2}$ in lower dimensional spaces. This tool is particularly useful for safety verification of large scale systems, since it allows the safety verification to be completed by analyzing its less complicated subsystems rather than analyzing the overall complex system.
\end{remark}

The small-gain condition (\ref{eq:small-gain}) is different from the traditional small-gain condition of \cite{jiang1996a}, whose analysis tools cannot be directly employed in our result. To prove Theorem \ref{thm:small-gain}, we first introduce two lemmas, which are extended from those given in \cite[Appendix]{jiang1996a}.

\begin{lemma}\label{eq:rho-construct}
Let $\rho_0:\R\rightarrow\R$ be a continuous function with $\rho_0(r)<0$ for all $r<0$ and $\rho_0(r)>0$ for all $r>0$. Then there exists a continuous function $\rho:\R\rightarrow\R$ such that
\begin{itemize}
  \item $\rho_0(r)<\rho(r)<0$ for all $r<0$, and $0<\rho(r)<\rho_0(r)$ for all $r>0$;
  \item $\rho$ is continuously differentiable on $\R$, and $\rho'(r)<\frac{1}{2}$ for all $r\in\R$.
\end{itemize}
\end{lemma}

\noindent
\textbf{Proof.}
See Appendix.
\hfill $\Box$

\begin{lemma}\label{eq:prelemma-small-gain-proof}
Let $\phi_i$ be of extended class $\K_\infty$ and satisfy (\ref{eq:small-gain}).
Then there exists an extended class $\K_\infty$ function $\phi$ such that
\begin{itemize}
  \item $\phi_1^{-1}(r)<\phi(r)<\phi_2(r)$ for all $r<0$, and $\phi_2(r)<\phi(r)<\phi_1^{-1}(r)$ for all $r>0$;
  \item $\phi(r)$ is continuously differentiable on $\R\backslash0$, and $\phi'(r)>0$ for all $r\in\R\backslash0$.
\end{itemize}
\end{lemma}
\noindent
\textbf{Proof.}
See Appendix.
\hfill $\Box$
\vskip5pt

Then it is time to prove Theorem \ref{thm:small-gain}.

\vskip5pt
\noindent
\textbf{Proof of Theorem \ref{thm:small-gain}.}
By applying Lemma \ref{eq:prelemma-small-gain-proof} to $\phi_1$ and $\phi_{2}$, one can select an extended class $\K_\infty$ function $\phi$, which is continuously differentiable on $\R\backslash0$, such that
\begin{subequations}\label{eq:phi2-sigma-hatphi1}
\begin{flalign}
    &\phi_1^{-1}(r)<\phi(r)<\phi_2(r),\;\;\forall r<0;
     \\
    &\phi_2(r)<\phi(r)<\phi_1^{-1}(r),\;\;\forall r>0.
\end{flalign}
\end{subequations}

Define a function
\begin{flalign}\label{eq:def-h}
    h(x)=\min\{\phi(h_1(x_1)),h_2(x_2)\}.
\end{flalign}
Let
\begin{flalign}
    &\setS_{\phi_1(\|h_2\|)}=\{x_1:h_1(x_1)+\phi_1(\|h_2\|)\geq0\},
     \nonumber\\
    &\setS_{\gamma_1(\|u_1\|)}=\{x_1:h_1(x_1)+\gamma_1(\|u_1\|)\geq0\},
     \nonumber\\
    &\setS_{\phi_2(\|h_1\|)}=\{x_2:h_2(x_2)+\phi_2(\|h_1\|)\geq0\},
     \nonumber\\
    &\setS_{\gamma_2(\|u_2\|)}=\{x_2:h_2(x_2)+\gamma_2(\|u_2\|)\geq0\}.
     \nonumber
\end{flalign}
Clearly, $\setS_1=\setS_{\phi_1(\|h_2\|)}\cup\setS_{\gamma_1(\|u_1\|)}$ and $\setS_2=\setS_{\phi_2(\|h_1\|)}\cup\setS_{\gamma_2(\|u_2\|)}$.

For every $(x_1,x_2)\in\partial\setS_{\phi_1(\|h_2\|)}\times\partial\setS_{\phi_2(\|h_1\|)}$, we have the following three cases.

\textbf{Case 1.} $\phi(h_1(x_1))<h_2(x_2)$.
Note that $h_i(x_i)\leq0$ when $(x_1,x_2)\in\partial\setS_{\phi_1(\|h_2\|)}\times\partial\setS_{\phi_2(\|h_1\|)}$. According to (\ref{eq:phi2-sigma-hatphi1}),
\begin{flalign}
    \phi_1^{-1}(h_1(x_1))<\phi(h_1(x_1))<h_2(x_2)
    \nonumber
\end{flalign}
and hence, it follows immediately from (\ref{eq:barrier-type-conditionA}) that
\begin{flalign}
    D_+{h}(x)
    &=\phi'(h_1(x_1))\nabla{h}_1(x_1)f_1(x_1,x_2,u_1)
     \nonumber\\
    &\geq-\phi'(h_1(x_1))\alpha_1(h_1(x_1))
    \label{eq:case1-dhB}
\end{flalign}
when $h(x)\leq-\hat{\gamma}_1(|u|)$ with $\hat{\gamma}_1(r)=-\phi(-\gamma_1(r))$.
The derivation of (\ref{eq:case1-dhB}) has employed the fact that $\phi'(h_1(x_1))>0$.

\textbf{Case 2.} $\phi(h_1(x_1))>h_2(x_2)$. According to  (\ref{eq:phi2-sigma-hatphi1}),
\begin{flalign}
    h_2(x_2)<\phi(h_1(x_1))<\phi_1(h_1(x_1))
    \nonumber
\end{flalign}
which implies
\begin{flalign}
    D_+{h}(x)
    =\nabla{h}_2(x_2)f_2(x_1,x_2,u_2)
    \geq-\alpha_2(h_2(x_2))
    \label{eq:case2-dhA}
\end{flalign}
when $h(x)\leq-\gamma_2(|u_2|)$.

\textbf{Case 3.} $\phi(h_1(x_1))=h_2(x_2)$. With Lemma \ref{lemma:dini}, it follows from (\ref{eq:case1-dhB}) and (\ref{eq:case2-dhA}) that
\begin{flalign}
    D_+{h}(x)
    \geq-\max\{\phi'(h_1(x_1))\alpha_1(h_1(x_1)), \alpha_2(h_2(x_2))\}
    \label{eq:case3-dhA}
\end{flalign}
when $h(x)\leq-\gamma(|u|)$ with $\gamma(r)=\hat{\gamma}_1(r)+\gamma_2(r)$.

Let $\alpha(r):=\max\{\phi'(r)\alpha_1(r), \alpha_2(r)\}$. Since $\phi'(r)>0$ for each $r\in\R$, $\alpha$ is an extended class $\K_\infty$ function. By summarizing the three cases above, we have
\begin{flalign}
    h(x)\leq-\gamma(|u|){\Rightarrow}D_+{h}(x)
    \geq-\alpha(h(x)).
    \label{eq:case3-dhA}
\end{flalign}

Without loss of generality, we can assume that $\alpha$ is smooth. In the following, we use the smoothing technique given in \cite[Remark 4.1]{lin1996smooth} to show that such an assumption is always possible.

Pick an extended class $\K_\infty$ function $\tilde{\alpha}$, which is smooth on $\R$, such that
\begin{flalign}
    &\tilde{\alpha}(s)\geq-{s}\alpha(s),\;\;\forall{s}\in[-1,0);
     \nonumber\\
    &\tilde{\alpha}(s)\leq{s}\alpha(s),\;\;\;\;\,\forall{s}\in(0,1].
     \nonumber
\end{flalign}
This is possible since $\alpha(r)$ is positive definite when $r>0$ and negative definite when $r<0$. Then let $\eta:\R\rightarrow\R$ be an extended class $\K_\infty$ function, which is smooth in $\R$, such that
\begin{itemize}
  \item $\eta(r)\leq{r}$ for all $r\in[-1,0)$;
  \item $\eta(r)<-\tilde{\alpha}(r)/\alpha(r)$ for all $r<-1$.
\end{itemize}
Define
\begin{flalign}
    \beta(r)=
    \left\{
      \begin{array}{ll}
        \int_0^s\eta(s), & \forall{r}\geq0; \\
        \int_s^0\eta(s), & \forall{r}<0.
      \end{array}
    \right.
\end{flalign}
Note that $\beta$ is of extended class $\K_\infty$. Let $W(h(x))=\beta(h(x))$ and then,
\begin{flalign}
    D_+W(h(x))
    =\beta'(h(x))D_+h(x)
    \geq\eta(h(x))\alpha(h(x))
\end{flalign}
when $h(x)\leq-\gamma(|u|)$. In the following, we show that $D_+W(h(x))$ is lower bounded by $-\tilde{\alpha}(h(x))$. Since $h(x)\leq-\gamma(|u|)\leq0$, we have the following two cases.

$\bullet$ If $h(x)\in[-1,0]$,
\begin{flalign}
    \eta(h(x))\leq{h}(x)\leq-\frac{\tilde{\alpha}(h(x))}{\alpha(h(x))}.
\end{flalign}

$\bullet$ If $h(x)\in(-\infty,-1)$, with the definition of $\eta$ on this interval, we have
\begin{flalign}
    \eta(h(x))\leq-\frac{\tilde{\alpha}(h(x))}{\alpha(h(x))}.
\end{flalign}
Therefore,
\begin{flalign}\label{eq:dw-final}
    h(x)\leq-\gamma(|u|)\Rightarrow{D}_+W(h(x))
    \geq-\tilde{\alpha}(h(x)).
\end{flalign}

By combining the proof of Theorem \ref{thm:ISSf} and Lemma \ref{lem:compar}, we can conclude from (\ref{eq:dw-final}) that the set $\setS_{\gamma(\|u\|)}=\{x\in\R^n:h(x)+\gamma(\|u\|)\geq0\}$ is forward robustly invariant. In other words, the interconnected system (\ref{eq:interconnect-system}) is ISSf on the set $\setS_{\gamma(\|u\|)}$.
\hfill $\Box$
\vskip5pt

Before the end of this section, we provide a simple example for illustration.

\vskip5pt
\noindent
\textbf{Example 1.}
Consider the interconnected system
\begin{flalign}
    &\dot{x}_1=-x_1-0.24x_2\sin(x_1-x_2)-0.5u_1^3,
     \nonumber\\
    &\dot{x}_2=-x_2-0.16x_1\sin(x_2-x_1)-0.5u_2.
     \nonumber
\end{flalign}
Suppose that $x_1$-subsystem and $x_2$-subsystem are ISSf on the safety sets $\setS_{1}$ and $\setS_{2}$, respectively, where
\begin{flalign}
    &\setS_{1}=\big\{x_1:x_1+\max\{0.96\|x_2\|,2\|u_1\|^3\}\geq0\big\},
     \nonumber\\
    &\setS_{2}=\big\{x_2:x_2+\max\{0.64\|x_1\|,2\|u_2\|\}\geq0\big\}.
     \nonumber
\end{flalign}
The objective of this example is to verify that the overall system is ISSf.

To achieve the control objective of Example 1, we choose $h_1(x_1)=x_1$ and $h_2(x_2)=x_2$, and then it follows that
\begin{flalign}
    &h_1(x_1)\leq-\max\{0.96|h_2|,2|u_1|^3\}\Rightarrow\dot{h}_1(x_1)\geq-0.5h_1(x_1),
     \nonumber\\
    &h_2(x_2)\leq-\max\{0.64|h_1|,2|u_2|\}\Rightarrow\dot{h}_2(x_2)\geq-0.5h_2(x_2).
     \nonumber
\end{flalign}
Thus, the barrier-type conditions given in (\ref{eq:barrier-type-conditionA}) and (\ref{eq:barrier-type-conditionB}) are satisfied. Moreover, we can specify that $\phi_1(r)=0.96r$, $\phi_2(r)=0.64r$, $\gamma_1(r)=2r^3$ and $\gamma_2(r)=2r$. Then we obtain $\phi_1\circ\phi_2(r)=0.6144r$, which implies that the small-gain condition (\ref{eq:small-gain}) is satisfied. By combining this with Lemma \ref{eq:prelemma-small-gain-proof}, we can take $\phi(r)=r$. With taking $h(x)=\min\{x_1,x_2\}$ and recalling (\ref{eq:case3-dhA}), we have
\begin{flalign}
    h(x)
    \leq-2|u|^3-2|u|
    {\Rightarrow}D_+{h}(x)
    \geq-0.5h(x),
    \nonumber
\end{flalign}
which implies the set
\begin{flalign}
    \setS=\{x:h(x)+2\|u\|^3+2\|u\|\geq0\}
    \nonumber
\end{flalign}
is forward robustly invariant. Clearly, we can see that $\setS$ is a safety set obtained from the safety sets $\setS_{1}$ and $\setS_{2}$ in the lower dimensional spaces.

\section{Conclusion}
We have developed a small-gain theorem based on ISSf-barrier functions for safety verification. It is shown that an interconnected system, which consists of two ISSf subsystems, is ISSf if the absolute value of the composition of ISSf gains  of two subsystems is smaller than that of the identity function.

\section*{Appendix}

\subsection{Proof of Lemma \ref{eq:rho-construct}}
Suppose $-\frac{1}{2}\leq\rho_0(r)<0$ for all $r<0$ and $0<\rho_0(r)\leq\frac{1}{2}$ for all $r>0$. Otherwise, we use
\begin{flalign}
    \left\{
      \begin{array}{ll}
        \max\{-\frac{1}{2},\rho_0(r)\}, & \text{if } r<0 \\
        0, & \text{if } r=0\\
        \min\{\frac{1}{2},\rho_0(r)\}, & \text{if } r>0
      \end{array}
    \right.
    \nonumber
\end{flalign}
to replace $\rho_0(r)$. Let $\rho_1(0)=0$ and
\begin{flalign}
    \rho_1(r)=\left\{
             \begin{array}{ll}
               \rho_1^-(r), & \text{if } r<0 \\
               \rho_1^+(r), & \text{if } r>0
             \end{array}
           \right.
\end{flalign}
with
\begin{flalign}
    &\rho_1^-(r)=\left\{
                  \begin{array}{ll}
                    \max_{s\in[-2,r]}\rho_0(s) & \text{if } -1\leq{r}<0; \\
                    \max_{s\in[r-1,-1]}\rho_0(s) & \text{if } r<-1;
                  \end{array}
                \right.
    \nonumber\\
    &\rho_1^+(r)=\left\{
                  \begin{array}{ll}
                    \min_{s\in[r,2]}\rho_0(s) & \text{if } 0<r\leq1 \\
                    \min_{s\in[1,r+1]}\rho_0(s) & \text{if } r>1.
                  \end{array}
                \right.
    \nonumber
\end{flalign}
Since $\rho_0(r)<0$ for $r\in[-2,0)$,
\begin{flalign}
    \lim_{r\rightarrow0^-}\rho_1(r)=\max_{s\in[-2,0]}\rho_0(s)=0;
    \nonumber
\end{flalign}
and analogously,
\begin{flalign}
    \lim_{r\rightarrow0^+}\rho_1(r)=\min_{s\in[r,2]}\rho_0(s)=0.
    \nonumber
\end{flalign}
Thus, $\rho_1$ is continuous at zero. Clearly, $\rho_1$ is continuous at $r=-1$ and at $r=1$. Thus, $\rho_1$ is continuous on $\R$.

To get a desired function $\rho$ satisfying the properties given in Lemma \ref{eq:rho-construct}, we take
\begin{flalign}\label{eq:rho}
    \rho(r)=
    \left\{
      \begin{array}{ll}
        \int_{r}^{r+1}\rho_1(s)\text{d}s, & \text{if } r<-1; \\
        \int_{r}^0\rho_1(s)\text{d}s, & \text{if } -1\leq{r}<0; \\
        \int_{0}^{r}\rho_1(s)\text{d}s, & \text{if } 0\leq{r}\leq1; \\
        \int_{r-1}^{r}\rho_1(s)\text{d}s & \text{if } r>1.
      \end{array}
    \right.
\end{flalign}
Since $\rho_1$ is a continuous function, we can verify that $\rho$ is continuously differentiable on $\R$. Note that $\rho_0(r)\leq\rho_1(r)<0$ for $r<0$, and $0<\rho_1(r)\leq\rho_0(r)$ for $r>0$. As a result, it is easy to see that $\rho'(r)\leq|\rho_1(r)|\leq\frac{1}{2}$ for all $r\in\R$. The second requirement given in Lemma \ref{eq:rho-construct} is met.

Next, we show that $\rho$ also meets the first requirement of Lemma \ref{eq:rho-construct}. Note that $\rho_1$ is not increasing on $(-\infty,-1)\cup(1,\infty)$, and not decreasing on $(-1,1)$. Also note that $\rho_1(r+1)\geq\rho_0(r)$ for $r\leq-1$, and $\rho_1(-1)\leq\rho_0(r)$ for all $r\geq1$. Then, with (\ref{eq:rho}), it can be observed that
\begin{flalign}
  \begin{array}{ll}
    \rho(r)\geq-{r}\rho_1(r)\geq\rho_0(r), & \forall{r}\in[-1,0); \\
    \rho(r)\geq\rho_1(-1)\geq\rho_0(r), & \forall{r}\in(-2,-1);\\
    \rho(r)\geq\rho_1(r+1)\geq\rho_0(r), & \forall{r}\in(-\infty,-2].\\
  \end{array}
   \nonumber
\end{flalign}
The first and the third inequalities are easy to verify by noting that $\rho_1$ is not decreasing on $(-1,0)$ and not increasing on $(-\infty,-2]$. For the second inequality, note that $\rho_1(-1)\leq\rho_1(r)<0$ for all $r\in(-2,-1)$. Similarly,
\begin{flalign}
  \begin{array}{ll}
    \rho(r)\leq{r}\rho_1(r)\leq\rho_0(r), & \forall r\in(0,1]; \\
    \rho(r)\leq\rho_1(1)\leq\rho_0(r), & \forall r\in(1,2);\\
    \rho(r)\leq\rho_1(r-1)\leq\rho_0(r), & \forall r\in[2,\infty).
  \end{array}
   \nonumber
\end{flalign}
Thus, the conclusion follows.

\subsection{Proof of Lemma \ref{eq:prelemma-small-gain-proof}}
Let
\begin{flalign}
    \rho_0(r)=\frac{1}{2}[r-\phi_1\circ\phi_2(r)].
\end{flalign}
According to (\ref{eq:small-gain}),
\begin{flalign}
    &\phi_1\circ\phi_2(r)>r-\rho_0(r),\;\;\text{if } r<0;
     \nonumber\\
    &\phi_1\circ\phi_2(r)<r-\rho_0(r),\;\;\text{if } r>0;
     \nonumber
\end{flalign}
and hence,
\begin{flalign}
    &\phi_2(r)>\phi_1^{-1}(r-\rho_0(r)),\;\;\forall r<0;
     \nonumber\\
    &\phi_2(r)<\phi_1^{-1}(r-\rho_0(r)),\;\;\forall r>0.
     \nonumber
\end{flalign}
By Lemma \ref{eq:rho-construct}, there exists a continuously differentiable function $\rho:\R\rightarrow\R$ with $\rho'(r)\leq\frac{1}{2}$ such that $\rho_0(r)<\rho(r)<0$ for each $r<0$, and $0<\rho(r)<\rho_0(r)$ for each $r>0$. Without loss of generality, we assume that $|\rho(r)|<|r|$. Now we let $\phi(0)=0$ and
\begin{flalign}
    \phi(r)=\frac{1}{\rho(r)}\int_{r-\rho(r)}^{r}\phi_1^{-1}(s)\text{d}s,\;\;\forall{r}\neq0.
\end{flalign}
With such a function, it yields immediately that $\phi_1^{-1}(r)<\phi(r)<\phi_1^{-1}(r-\rho(r))<\phi_2(r)$ for all $r<0$, and $\phi_2(r)<\phi_1^{-1}(r-\rho(r))<\phi(r)<\phi_1^{-1}(r)$ for all $r>0$. Recall that $\phi_1^{-1}(0)=0$ and $\phi_1^{-1}$ is continuous on $\R$. It means that $\lim_{r\rightarrow0^+}\phi(r)=\lim_{r\rightarrow0^-}\phi(r)=0$, which further implies that $\phi(r)$ is continuous at zero, and consequently, $\phi(r)$ is continuous on $\R$ as well.

Since $\rho$ is continuously differentiable on $\R$, $\phi$ is continuously differentiable on $\R\backslash0$, and
\begin{flalign}\label{eq:dsigmaA}
    \phi'(r)
    &=-\frac{\rho'(r)}{\rho^2(r)}\int_{r-\rho(r)}^{r}\phi_1^{-1}(s)\text{d}s
     \nonumber\\
    &\;\;\;\;
      +\frac{1}{\rho(r)}[\phi_1^{-1}(r)-\phi_1^{-1}(r-\rho(r))(1-\rho'(r))]
     \nonumber\\
    &=\frac{1}{\rho(r)}\bigg[\phi_1^{-1}(r)-\phi_1^{-1}(r-\rho(r))-\frac{\rho'(r)}{\rho(r)}
     \nonumber\\
    &\;\;\;\;\;\;\;\;\;\;\;\;
      \times\int_{r-\rho(r)}^{r}\phi_1^{-1}(s)\text{d}s+\rho'(r)\phi_1^{-1}(r-\rho(r))\bigg].
\end{flalign}
Then we show that $\phi'(r)>0$ for all $r<0$. According to $\rho'(r)$, we have the following two cases.

$\bullet$ If $\rho'(r)\leq0$, then
\begin{flalign}
    &-\frac{\rho'(r)}{\rho(r)}\int_{r-\rho(r)}^{r}\phi_1^{-1}(s)\text{d}s
     +\rho'(r)\phi_1^{-1}(r-\rho(r))
     \nonumber\\
    &\;\;\;\;
     \leq-\rho'(r)\phi_1^{-1}(r-\rho(r))+\rho'(r)\phi_1^{-1}(r-\rho(r))
     \nonumber\\
    &\;\;\;\;
     \leq0
\end{flalign}
and hence, by substituting this into (\ref{eq:dsigmaA}), one obtains $\phi'(r)>0$.

$\bullet$ If $\rho'(r)>0$, then
\begin{flalign}
    &-\frac{\rho'(r)}{\rho(r)}\int_{r-\rho(r)}^{r}\phi_1^{-1}(s)\text{d}s
     +\rho'(r)\phi_1^{-1}(r-\rho(r))
     \nonumber\\
    &\;\;\;\;
     \leq-\rho'(r)\phi_1^{-1}(r)+\rho'(r)\phi_1^{-1}(r-\rho(r))
\end{flalign}
and hence,
\begin{flalign}
    \phi'(r)
    &\geq\frac{1}{\rho(r)}[1-\rho'(r)][\phi_1^{-1}(r)-\phi_1^{-1}(r-\rho(r))]
     \nonumber\\
    &\geq\frac{1}{2\rho(r)}[\phi_1^{-1}(r)-\phi_1^{-1}(r-\rho(r))]
     \nonumber\\
    &>0.
\end{flalign}
By combining the two cases above, one can conclude that $\phi'(r)>0$ for all $r<0$. Moreover, we can obtain from \cite[Lemma A.1]{jiang1996a} that $\phi'(r)>0$ for all $r>0$.


\ifCLASSOPTIONcaptionsoff
  \newpage
\fi


\end{CJK}
\end{document}